\documentclass{PoS}
\usepackage{amsmath}

\title{Threshold resummation for Drell-Yan production: theory and phenomenology}

\ShortTitle{Threshold resummation for Drell-Yan production: theory and phenomenology}

\author{\speaker{Marco Bonvini}%
        \\
        Dipartimento di Fisica, Universit\`a di Genova and
        INFN, Sezione di Genova, Italy\\
        E-mail: \email{marco.bonvini@ge.infn.it}}

\abstract{%
We present a phenomenological study of Drell-Yan pair production at hadron colliders
based on the NNLO fixed order calculation and on NNLL resummation of threshold logarithms.
We give an argument to prove that resummation effects are relevant
also for values of $x=M^2/s$ far from threshold.
We compare different prescriptions for the calculation of resummed quantities,
emphasizing the differences coming from subleading terms, which are important
when $x$ is small. We present phenomenological predictions for Drell-Yan rapidity
distributions at the LHC, we study the ambiguity related to the resummation prescription,
and we compare it to that coming from scale variation.
}

\FullConference{XVIII International Workshop on Deep-Inelastic Scattering and Related Subjects\\
		 April 19 -23, 2010\\
		 Convitto della Calza, Firenze, Italy}

\newcommand\as{\alpha_{\rm s}}
\def\beq{\begin{equation}}
\def\eeq{\end{equation}}
\def\({\left(}
\def\){\right)}
\newcommand{\plus}[1]{\left[ #1 \right]_+}
\def\ab{\bar\alpha}

\newcommand{\cteq}{CTEQ6.6}

\begin{document}

\section{Introduction: threshold resummation in the Drell-Yan process}

A generic (differential) parton model cross section $\sigma$ at hadron colliders
can be written as a convolution
\beq\label{eq:cs-hadr}
\sigma(x) = \int_x^1 \frac{dz}{z} \, L\(\frac{x}{z}\)\, \hat\sigma(z)
\eeq
where $x=M^2/s$ ($M$ is the Drell-Yan pair invariant mass, $\sqrt{s}$ the hadronic cms energy),
$L(z)$ is a parton luminosity and $\hat\sigma(z)$ is the parton-level cross-section. 
In $\hat\sigma(z)$ logarithms of $1-z$ appear at all orders in $\as$
\beq\label{eq:logs}
\as^n \,\plus{\frac{\log^{l}(1-z)}{1-z}}
\;, \qquad
l = 2n-1,\ldots,0 \;,
\eeq
and in the partonic threshold limit $z\to 1$ these logarithms become large,
spoiling the reliability of the perturbative expansion.
All order resummation \cite{res} is needed in this case.

\section{Relevance of resummation of $\log(1-z)$ at small $x$}
\label{sec:saddle}

\begin{figure}[b]
\begin{center}
\includegraphics[width=0.7\textwidth]{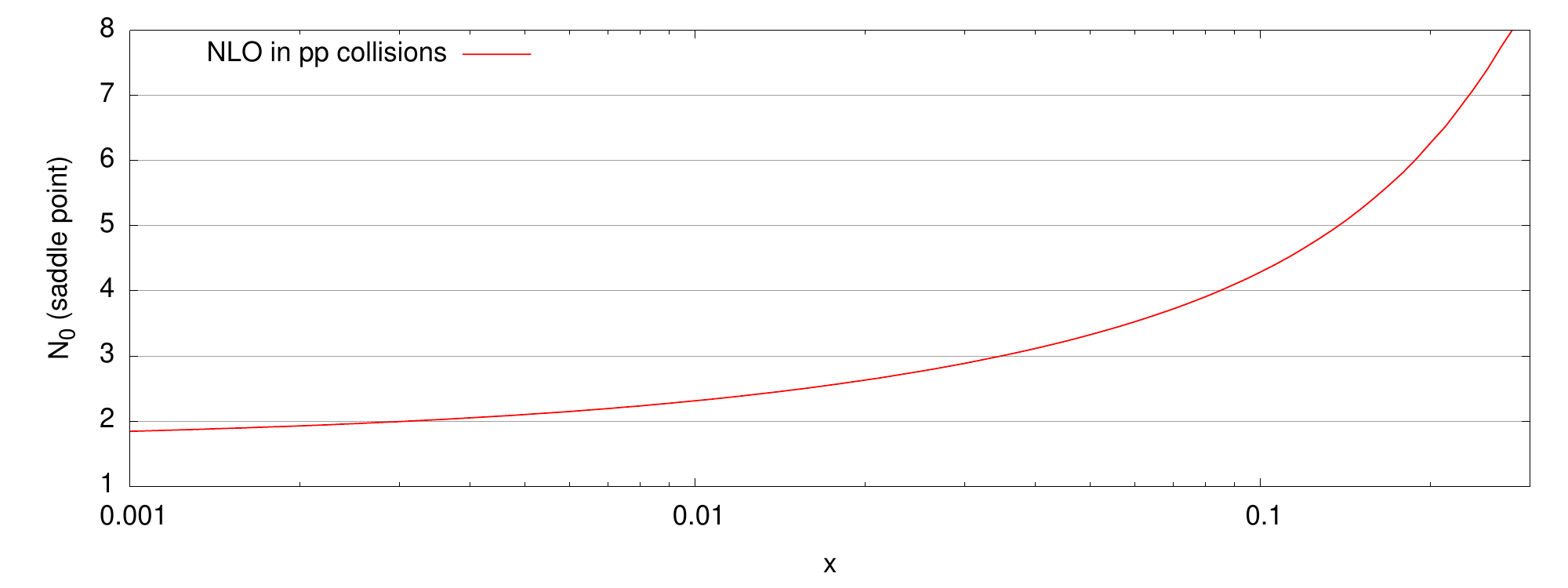}
\caption{Position of the saddle point $N_0$ as a function of $x$
for the order-$\as$ Drell-Yan $q\bar q$ cross section.}
\label{fig:saddle}
\end{center}
\end{figure}

\begin{figure}[htb]
\begin{center}
\includegraphics[width=0.65\textwidth]{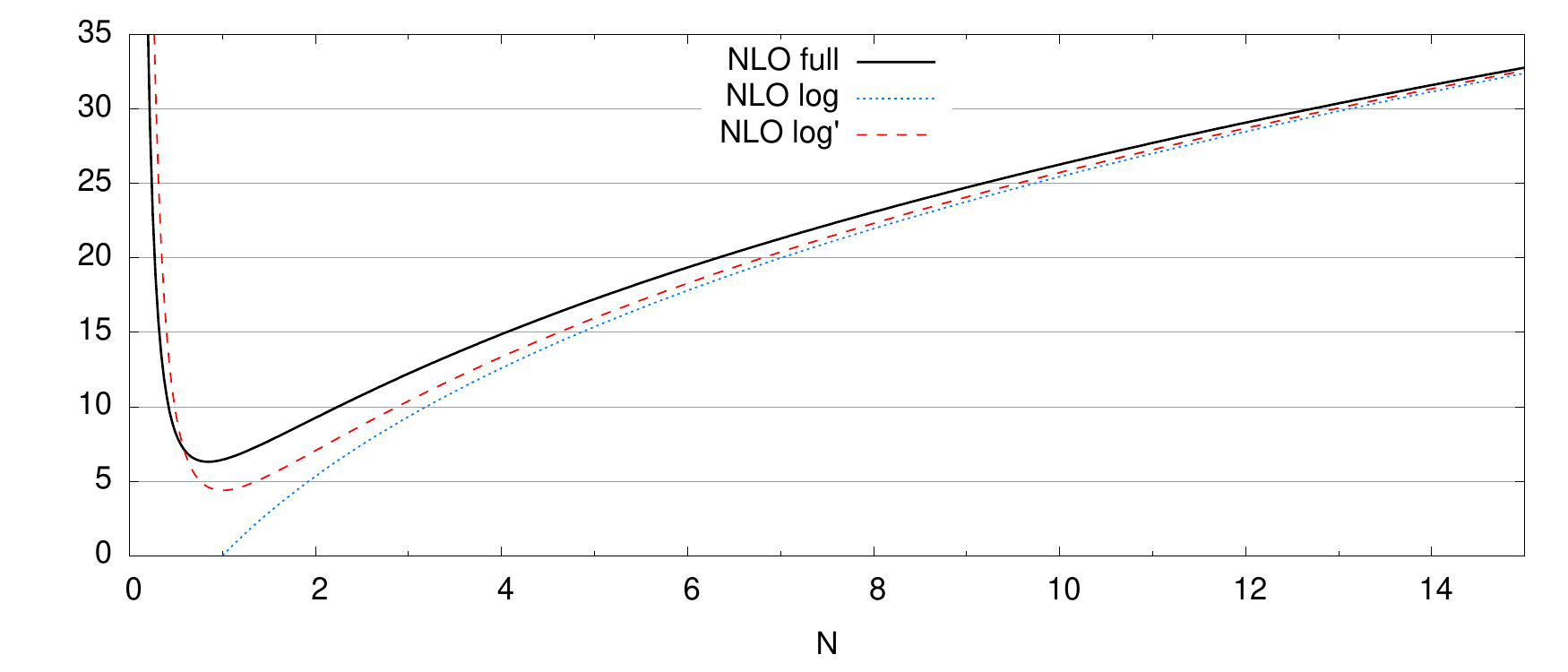}
\caption{NLO Drell-Yan coefficient
function as a function of $N$, and its logarithmic approximations.}
\label{fig:NLO}
\end{center}
\end{figure}

We see from eq.~\eqref{eq:cs-hadr} that the partonic threshold region $z\to 1$ is always
included in the convolution integral. To understand when resummation is relevant at the hadron level
we need to establish when $z\to 1$ region gives the dominant contribution to the integral.
Qualitatively \cite{catani} this happens when
the shape of the parton luminosity enhances that region of integration.
A quantitative answer \cite{bfr2} can be given working in $N$-space, where $N$ is the Mellin-conjugate
variable to $x$.
Eq.~\eqref{eq:cs-hadr} can be rewritten as a Mellin inversion integral
\beq\label{eq:sigma_hadr}
\sigma(x) = \frac{1}{2\pi i}\int_{c-i\infty}^{c+i\infty} dN \; x^{-N} \, L(N) \, \hat\sigma(N)
=\frac{1}{2\pi i}\int_{c-i\infty}^{c+i\infty} dN \; e^{N\log\frac{1}{x} + \log L(N) + \log \hat\sigma(N)}
\eeq
where $L(N)$, $\hat\sigma(N)$ are the Mellin transforms of $L(z)$, $\hat\sigma(z)$. This
inversion integral is dominated by the region where the exponent has a minimum in the positive
real axis (saddle point $N_0$).
By a general argument, one can show \cite{bfr2} that a saddle point always exists.
We show in Fig.~\ref{fig:saddle} the position of the saddle point $N_0$ as a function of $x$,
where $\hat\sigma(N)$ is the DY $q\bar q$ contribution at NLO
and the parton luminosity is computed using \cteq{} in $pp$ collisions.
The saddle point $N_0$ is monotonically increasing with $x$, meaning that
at large $x$ the contribution to the cross section mainly comes from
the large $N$ region, which corresponds to the partonic threshold region.
For smaller $x$, the saddle point $N_0$ decreases; to evaluate
when the large logarithms still give the dominant contribution,
we use a fixed order computation to compare the logarithmic terms with the full result.
The DY $q\bar q$ contribution at the NLO
is given by (up to an electroweak normalization factor)
\begin{align}
&\hat\sigma(z)=\delta(1-z)+\frac{\as}{\pi}C_1(z) + O(\as^2)
\\
&C_1(z) = 4C_F\left\{\plus{\frac{\log(1-z)}{1-z}}
-\frac{\log\sqrt{z}}{1-z} - \frac{(1+z)}{2}\log\frac{1-z}{\sqrt{z}}
+\(\frac{\pi^2}{12} -1\)\delta(1-z)
\right\}\;.\label{eq:C1}
\end{align}
In Fig.~\ref{fig:NLO} we show the Mellin transform of $C_1(z)$ (black solid curve)
and its logarithmic part, the first term in eq.~\eqref{eq:C1} (blue dotted curve).
The red dashed curve corresponds to the first two terms in eq.~\eqref{eq:C1}:
indeed, the term $\frac{\log\sqrt{z}}{1-z}$ has the same dynamical origin of the
logarithm inside the plus distribution, and may be included as well in resummation (see Sect.~\ref{sec:res}).
Inspection of Fig.~\ref{fig:NLO} leads to the conclusion that,
down to values of $N$ around $2$, the logarithmic part dominates
the partonic NLO cross section.
Hence, if the saddle point $N_0 \gtrsim 2$, the integrand defining
the hadronic cross section $\sigma(x)$ in eq.~\eqref{eq:sigma_hadr}
is dominated by the logarithmic part of the partonic cross section.
Comparing with Fig.~\ref{fig:saddle} we can conclude that
for values of $x\gtrsim 0.003$ resummation of the logarithms is relevant,
in the sense that it includes to all orders the dominant terms of the
perturbative coefficients.
Note that this value is rather small, and usually considered too small
for threshold resummation to have a sizable effect.

\section{Comparison between the Borel prescription and the minimal prescription}
\label{sec:res}

Threshold resummation is performed in $N$ space~\cite{res}, and the generic
resummed quantity can be written as a function of $\log\frac{1}{N}$
($h_k$ are coefficients which depend on $\as$)
\beq\label{eq:sigma_res}
\hat\sigma^{\rm res}(N) \equiv \Sigma\(\ab\log\frac{1}{N}\) = \sum_{k=1}^\infty h_k\, \ab^k\,\log^k\frac{1}{N}
\;,\qquad
\ab = 2\beta_0\as \;.
\eeq
This series has a radius of convergence $1$ in $\ab\log\frac{1}{N}$, since the sum
has a branch cut in $e^{\frac{1}{\ab}}\leq N\leq +\infty$ due to the Landau pole of $\as$.
For this reason, the Mellin inversion integral \eqref{eq:sigma_hadr} is not defined,
because the parameter $c$ should be to the right of all the singularities of the integrand,
but the cut makes this condition impossible to be satisfied.
Otherwise stated, the series obtained performing the Mellin inversion of \eqref{eq:sigma_res}
term by term is divergent \cite{frru_afr,bfr2}.
Hence a prescription is needed to obtain a resummed cross section in the physical $x$ space.

In \cite{cmnt} the minimal prescription has been proposed: the idea is simply to
choose the integral path in \eqref{eq:sigma_hadr} to the left of the cut.
This choice has some good properties, one of them being that the result is asymptotic
to the divergent series.
However, it also has some undesired features: first it is defined only at the hadron level,
and the integral gets a contribution from an unphysical region of the parton densities~\cite{cmnt};
second, there is no way to control subleading terms.
To clarify this second statement, we apply the minimal prescription to
the $k$-th term of the series \eqref{eq:sigma_res}, i.e.~$\log^k\frac{1}{N}$.
In this case the integral is an exact Mellin inversion, and the result is a sum
of terms of the form
\beq
\plus{\frac{\log^j\log\frac{1}{z}}{\log\frac{1}{z}}}\;,
\qquad j = k-1, \ldots , 0 \;,
\eeq
which are an approximation of the physical logarithms \eqref{eq:logs} in the limit $z\to 1$.
There is no way to force the minimal prescription to reproduce the physical logarithms.

In \cite{frru_afr} a different prescription, based on the Borel summation of
the divergent series \eqref{eq:sigma_res}, has been proposed.
We present here a simpler proof of the Borel prescription formula, following~\cite{bfr2}.
First, we rewrite the Mellin inversion of $\log^k\frac{1}{N}$ as
\beq\label{eq:mellin_logk}
\frac{1}{2\pi i}
\int_{c-i\infty}^{c+i\infty} dN\,z^{-N}\log^k\frac{1}{N} 
=
\frac{k!}{2\pi i}\oint\frac{d\xi}{\xi^{k+1}}
\,\frac{1}{\Gamma(\xi)}\plus{\log^{\xi-1}\frac{1}{z}}
\eeq
where the integral path in the r.h.s.~is any closed contour which contains $\xi=0$.
Next, we use the identity
\beq
k! = \int_0^{+\infty} dw\,e^{-w}\,w^k
\eeq
to eliminate the $k!$ in \eqref{eq:mellin_logk}. The Borel method consists
in exchanging the $w$ integral with the series in \eqref{eq:sigma_res}, obtaining
\beq
\hat\sigma^{\rm res}(z) =
\frac{1}{2\pi i} \oint\frac{d\xi}{\xi \Gamma(\xi)}
\int_0^{+\infty} \frac{dw}{\ab}\,e^{-w/\ab}\;
\Sigma\(\frac{w}{\xi}\)
\, \plus{\log^{\xi-1}\frac{1}{z}} \;.
\eeq
However, this expression is still divergent.
The Borel prescription is obtained putting an upper cutoff $C$ to the $w$ integral;
this choice amounts to the inclusion of an higher twist term.
It is shown in \cite{frru_afr} that this result is asymptotic to the divergent series.
Most importantly, the $z$ dependence is explicit, and can be modified
in order to better reproduce the physical logarithms.
Since in the previous section we have seen that the inclusion of the $\log\sqrt{z}$ term
can help to better approximate the shape of the full fixed order, we include
also these logs and obtain our final expression:
\beq
\hat\sigma^{\rm res}_{\rm BP}(z) =
\frac{1}{2\pi i} \oint\frac{d\xi}{\xi \Gamma(\xi)}
\int_0^{C} \frac{dw}{\ab}\,e^{-w/\ab}\;
\Sigma\(\frac{w}{\xi}\)
\, \plus{(1-z)^{\xi-1}}z^{-\xi/2} \;.
\eeq
The main features of the Borel prescription 
are that it is a parton level expression and that we can control the $z$ dependence.
Furthermore, the parameter $C$ can be used to estimate the ambiguity of the prescription.

\section{Results: Drell-Yan rapidity distributions at hadron colliders}

\begin{figure}[t]
\begin{center}
\includegraphics[width=0.495\textwidth,page=1]{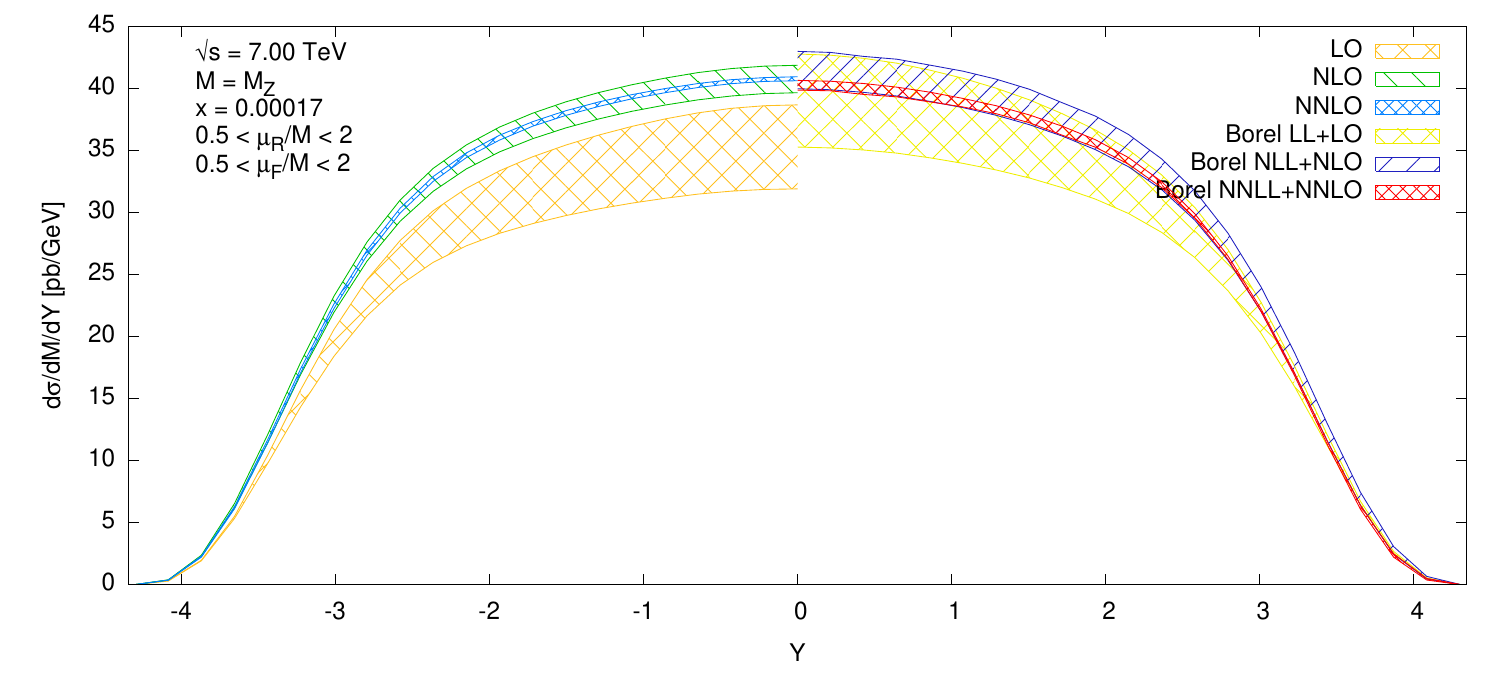}
\includegraphics[width=0.495\textwidth,page=2]{proc_graph_rap_pp_7_Z}
\includegraphics[width=0.495\textwidth,page=1]{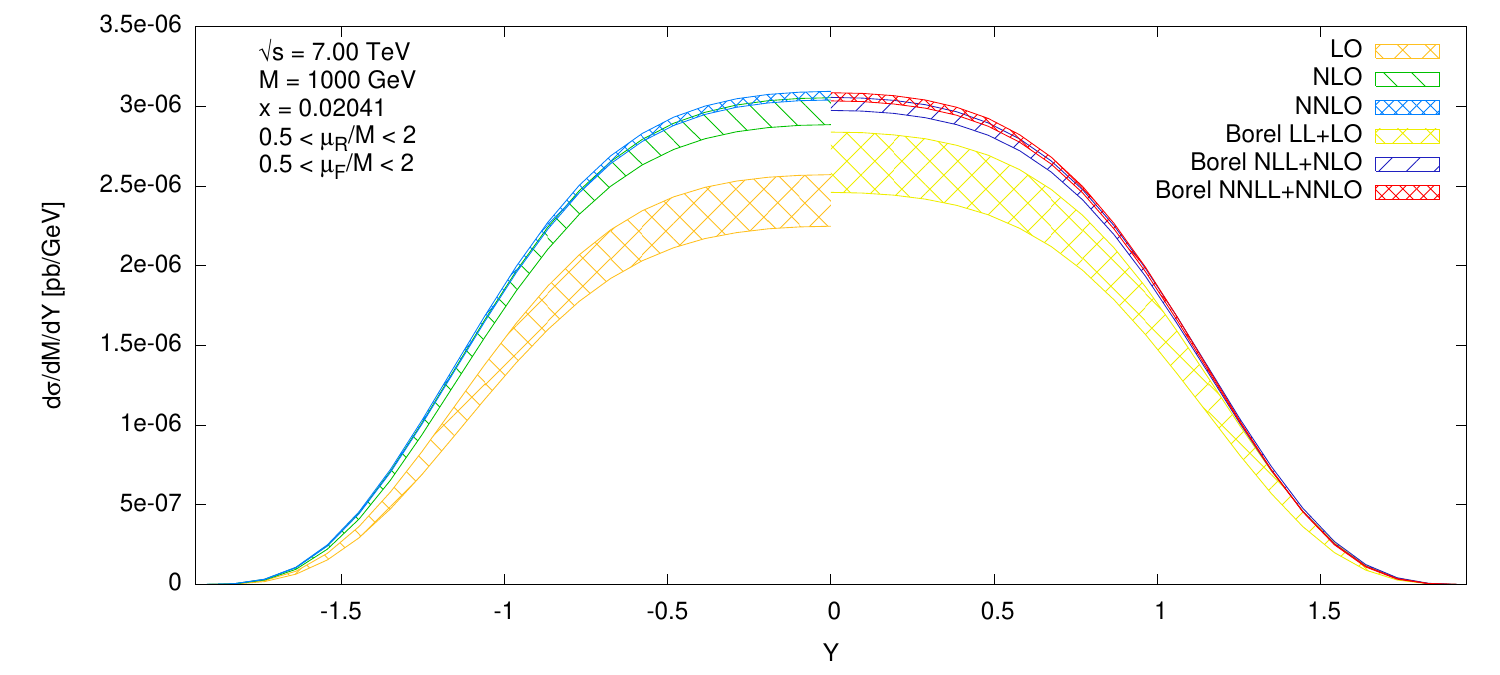}
\includegraphics[width=0.495\textwidth,page=2]{proc_graph_rap_pp_7_1000}
\caption{Rapidity distribution of neutral Drell-Yan pairs
of invariant mass $M=m_Z$ (upper plots) and
$M=1$~TeV (lower plots) in $pp$ collisions at $\sqrt{s}=7$~TeV (using \cteq\ pdf set and $\as(m_Z) = 0.118$).}
\label{fig:rapLHC7}
\end{center}
\end{figure}

In Fig.~\ref{fig:rapLHC7} we show the theoretical predictions for the production of a neutral DY pair
in $pp$ collisions at $\sqrt{s}=7$~TeV.
In all plots, the $Y<0$ region shows the fixed order results, at LO, NLO and NNLO
accuracy. The $Y>0$ region includes the resummed results (using $C=2$ Borel prescription in the left plots
and minimal prescription in the right plots) at LL, NLL, NNLL accuracy.
The error bands are obtained by varying the renormalization and factorization
scales independently by a fator of $2$ around $M$.
The pdf uncertainty is not included.

The upper plots refer to the production of a DY pair of invariant mass $M=m_Z$.
In this case $x\sim 10^{-4}$ is rather small (with respect to the value $0.003$ found in Sect.~\ref{sec:saddle}),
meaning that the large $z$ region of the partonic cross section does not give the dominant contribution
to the cross section.
Indeed, there is a quite large difference between the Borel prescription and the minimal
prescription results, coming from the different way of including subleading terms.
We believe that this discrepancy can be used in order to better estimate the ambiguity
due to the unknown higher orders. 
Moreover, the resummed contribution does not reduce the scale dependence, since resummation
only affects the $q\bar q$ channel, and the reduction of scale dependence is
a combined effect of the $q\bar q$, $qg$ and $gg$ contributions.

The lower plots refer to the production of a DY pair of invariant mass $M=1$~TeV.
In this case, $x\sim 0.02$ is in the region for which resummation is relevant.
At the NNLO+NNLL level of accuracy, the inclusion of the resummed term gives a small improvement:
the scale uncertainty is mildly reduced.

\section{Conclusions}

In conclusion, we have shown by a quantitative argument that inclusion
of threshold logarithms at all orders does improve the accuracy of QCD perturbative predictions
even when $x=M^2/s$ is not close to $1$ (down to $x\sim 10^{-3}-10^{-2}$).
Furthermore, we have presented a realistic phenomenological application
of the Borel prescription for the computation of resummed cross sections.


\begin{thebibliography}{99}
  \bibitem{res}
    S.~Catani and L.~Trentadue,
    Nucl.\ Phys.\  B {\bf 327} (1989) 323;
    G.~Sterman,
    Nucl.\ Phys.\  B {\bf 281} (1987) 310;
    S.~Forte and G.~Ridolfi,
    Nucl.\ Phys.\  B {\bf 650} (2003) 229
    [arXiv:hep-ph/0209154].
  \bibitem{catani}
    S.~Catani, D.~de Florian and M.~Grazzini,
    JHEP {\bf 0201} (2002) 015
    [arXiv:hep-ph/0111164].
  \bibitem{bfr2}
    M.~Bonvini, S.~Forte and G.~Ridolfi,
    arXiv:1009.5691 [hep-ph].
  \bibitem{frru_afr}
    S.~Forte, G.~Ridolfi, J.~Rojo and M.~Ubiali,
    Phys.\ Lett.\  B {\bf 635} (2006) 313
    [arXiv:hep-ph/0601048];
    R.~Abbate, S.~Forte and G.~Ridolfi,
    Phys.\ Lett.\  B {\bf 657} (2007) 55
    [arXiv:0707.2452 [hep-ph]].
  \bibitem{cmnt}
    S.~Catani, M.~L.~Mangano, P.~Nason and L.~Trentadue,
    Nucl.\ Phys.\  B {\bf 478} (1996) 273
    [arXiv:hep-ph/9604351].

\end{thebibliography}
\end{document}